\documentclass[aps,twocolumn, floats,preprintnumbers,showpacs,prd]{revtex4}
\usepackage{graphicx}
\usepackage{url}
\usepackage{amsmath}
\usepackage{amsfonts}
\usepackage{amssymb}
\def\beq{\begin{equation}}
\def\eeq{\end{equation}}
\def\beqa{\begin{eqnarray}}
\def\eeqa{\end{eqnarray}}
\def\bea{\begin{eqnarray}}
\def\eea{\end{eqnarray}}

\def\ltap{\ \raise.3ex\hbox{$<$\kern-.75em\lower1ex\hbox{$\sim$}}\ }
\def\gtap{\ \raise.3ex\hbox{$>$\kern-.75em\lower1ex\hbox{$\sim$}}\ }

\begin{document}
\preprint{SLAC-PUB-14617}
\preprint{UCI-TR-2011-21}
\title{
Dark Matter Jets at the LHC
}

\author{Yang Bai$^{a}$ and Arvind Rajaraman$^{b}$
\\
\vspace{2mm}
${}^{a}$SLAC National Accelerator Laboratory, 2575 Sand Hill Road, Menlo Park, CA 94025, USA, \\
${}^{b}$Department of Physics and Astronomy,
University of California, Irvine, CA 92697, USA
}
%\vspace{2mm}
%\date{\today}

\pacs{12.60.-i, 95.35.+d, 14.80.-j}

\begin{abstract}
We argue that  dark matter particles which have strong interactions
with  the Standard Model particles are not excluded by current astrophysical
constraints.  These dark matter particles have unique signatures at
colliders; instead of missing energy, the dark matter particles produce
jets. We propose a new search strategy for such strongly interacting particles
by looking for a signal of two trackless jets. We  show that suitable cuts can plausibly
allow us to find these signals at the LHC even in early data.
\end{abstract}
\maketitle
%\keywords{}

%%%%%%%%%%%%%%%%%%%%%%%%%%%%%%%%
%\section{Introduction}
%\label{sec:intro}
{\it{\textbf{Introduction.}}}
%%%%%%%%%%%%%%%%%%%%%%%%%%%%%%%
Many different astrophysical and cosmological measurements have definitively established the existence and abundance of dark matter (DM)~\cite{Komatsu:2010fb}. However, these observations do little to establish the masses and interaction strengths of these DM particles. {\it A priori}, the DM particles only need to have gravitational interactions; however, much stronger interaction strengths are not excluded. Similarly, DM models propose masses ranging from sub-ev~\cite{Duffy:2009ig} to several TeV~\cite{Bertone:2004pz}.

Perhaps the most popular candidate for DM is a weakly interacting massive particle (WIMP), interacting via  weak interactions with the Standard Model (SM) particles~\cite{Bertone:2004pz}. Many models of physics beyond the SM (such as supersymmetry and technicolor) naturally have stable WIMP candidates for DM. Furthermore, WIMPs of a mass close to a few hundred GeV naturally obtain the correct observed relic density through thermal interactions. Motivated by these considerations, many experiments have searched for WIMPs through direct detection of WIMP scattering  off nuclei. However, no unambiguous evidence of DM has been found so far~\cite{Ahmed:2009zw, Aprile:2010um}.

WIMPs can also be searched for at colliders like the Tevatron and the Large Hadron Collider (LHC). Because of its weakly-interacting strength with ordinary matter, the WIMPs escape
the detector and are undetected. This can lead to events where the visible particles have a
nonzero total transverse momentum (balanced by the escaping DM particles). Both the LHC and Tevatron have searched for such events with a large missing transverse momentum, and have seen no deviations from the predictions of the SM~\cite{Aaltonen:2008hh,Atlasmonojet} so far. This has placed significant constraints on many beyond-the-standard models with stable WIMP particles~\cite{Goodman:2010yf,Bai:2010hh,Goodman:2010ku}.

In view of these constraints, it may be worthwhile to reconsider alternative models of DM. It
is particularly interesting to consider models where the DM is 
a strongly interacting massive particle (SIMP)~\cite{Starkman:1990nj}.
Indeed, DM with strong interactions has features which may
enable it to be undetected in searches for WIMPs. For example, SIMPs may be stopped in the
atmosphere and earth, and cannot reach the detectors hundreds of meters underground. It is
therefore worthwhile to see if some portion of SIMP parameter space is still viable.

If the SIMPs have masses below a TeV, the Tevatron and LHC can copiously produce them in pairs (if their cosmological stability is protected by a ${\cal Z}_2$ symmetry). Unlike WIMPs, SIMPs could behave like a neutron and deposit energy or even stop in the calorimeters. Therefore, for a wide range of parameter space, SIMPs produce two jets without missing energy, and the current DM searches with missing energy are insensitive to SIMPs. Compared to the QCD jets, the jets from SIMPs have zero tracks and less electromagnetic activity in the electromagnetic calorimeter and therefore can be distinguished from the QCD dijet backgrounds. We will call these jets ``{\emph{dark matter jets}}".

We will begin by considering various constraints which have already been placed on
the SIMP parameter space. These come from many different varieties
of astrophysical constraints, and we review these in the following section.
We find that there are significant regions of parameter space which
are still allowed. We then discuss the strategies and the prospects of probing these
regions of parameter space at the LHC.

%%%%%%%%%%%%%%%%%%%%%%%%%%%%%%%
{\it{\textbf{Constraints on SIMPs.}}}
%%%%%%%%%%%%%%%%%%%%%%%%%%%%%%%
For the purposes of discussing constraints on SIMP models, it will often be
convenient to consider a model where the SM is extended by including a single
DM particle $\chi$ with a mass $m_\chi$. Another important parameter is the
total cross section of DM scattering off nucleons $\sigma_{\chi p}$. Various
constraints can be applied to the $m_\chi$--$\sigma_{\chi p}$ plane. We now
consider the most important constraints in turn.

\begin{figure}[!]
\begin{center}
\hspace*{-0.75cm}
\includegraphics[width=0.48\textwidth]{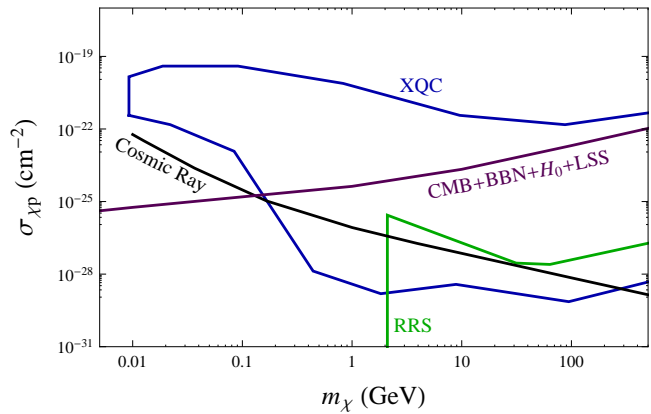}
\caption{Current experimental limits on SIMPs (see text for details). The regions above the black and purple solid lines and the regions enclosed by the blue and green contours are excluded.}
\label{fig:constraint}
\end{center}
\end{figure}

{\it Direct  Detection Searches}:
Some of the most robust bounds on the SIMP parameter space come from direct detection
searches. Ground-based experiments like CDMS and XENON place stringent bounds at somewhat lower cross sections. At higher cross sections, as already mentioned, the DM particles are stopped by the atmosphere and the ground-based experiments become insensitive to them.
However, high-altitude detectors (balloon-, satellite- or rocket-based experiments) can
search for interactions above the atmosphere, and these eliminate large swaths of the
parameter space. In particular, the RRS (balloon-based silicon semiconductor
detector)~\cite{Rich:1987st} and X-ray Quantum Calorimetry (XQC)~\cite{Wandelt:2000ad,Erickcek:2007jv} experiments exclude SIMPs over a large region of parameter space. We show the excluded regions at 90\% C.L. in the enclosed regions (the blue and green lines) of Fig.~\ref{fig:constraint}.

{\it Earth Heating Bounds}: A particularly interesting
bound comes from constraints on heat emitted from the Earth's core.
SIMPs can be captured by the Earth and accumulate over time at the Earth's core.  If SIMPs can self-annihilate into SM particles, the Earth's heat flow, which can be measured by placing detectors in deep underground shafts, would be modified. Mack and Beacom~\cite{Mack:2007xj} showed that this constraint effectively closes all of SIMP parameter space. However, this bound does not apply if SIMP self-annihilation is absent; for example asymmetric SIMP DM would evade this bound~\cite{Chivukula:1989qb}.

{\it Neutron Stars and Black Holes}: Another interesting bound can be imposed for scalar DM; these can be collected in the cores of neutron stars and cause neutron stars to collapse to black holes. This excludes scalar SIMPs to very low masses~\cite{Kouvaris:2011fi,McDermott:2011jp}. However, if the DM is fermionic, then this constraint is avoided.

{\it Cosmic Rays}: Another important constraint comes from cosmic rays~\cite{Cyburt:2002uw}. Protons in cosmic rays can scatter off DM particles and produce neutral pions, which would decay to photons and be visible in gamma ray telescopes. Using this fact, the authors of \cite{Cyburt:2002uw} placed a bound on DM-nucleon interactions.
However, these constraints depend on many factors; most importantly they assume
a form of the DM density near the galactic core. Since we are now dealing with a nonstandard form of DM with relatively strong interaction with baryons, these
densities may be significantly modified, making the cosmic ray constraints somewhat uncertain. In any case, we show the upper exclusion limit in the black line of Fig.~\ref{fig:constraint} by requiring the cosmic ray contributions from DM to not  exceed the observed flux.

{\it CMB and LSS}: The large scattering interaction between DM and baryons can also change the 
cosmic microwave background (CMB) anisotropy and large scale
structure (LSS) power spectrum. The actual constraints depend on the coherent or incoherent 
properties of the scattering. Taking the more stringent limits for a coherent 
scattering ($\sigma_{\chi He} = 16 \sigma_{\chi p}$), the upper constraints at 95\% C.L. are shown in the purple line of Fig.~\ref{fig:constraint}~\cite{Chen:2002yh}.

%{\it Milky Way Disruption}: For very large interaction cross sections, the dark matter
%is stopped by the Milky Way itself, and even the rocket borne experiments become
%insensitive to dark matter. However, for these large cross sections, the Milky Way disk
%would be disrupted by dark matter scattering, and this excludes such
%interactions~\cite{Starkman:1990nj,Natarajan:2002cw}. The constrains from Milky way disruption are %weaker than from cosmic rays and CMB plus LSS and are not shown in Fig.~\ref{fig:constraint}.

{\it Bound States}:
If SIMPs can form bound states with nucleons, then these states can be found in heavy water searches. These constraints are very severe, and would exclude such models. We must make sure that such bound states do not form.

The simplest way to make this happen would be if the force between DM and nucleons is repulsive.
This could happen if the force mediator was a vector, and the DM and SM particles all had
the same sign of charge. In that case, however, the charge neutrality of the
universe is violated, and while this is not obviously impossible, it
is difficult to find models where a nonzero gauge charge can be generated.

The alternative is to take the mediator to be a scalar and the
nucleons and DM to have opposite signs of the couplings.
%This indeed
%generates a repulsive force between DM and nucleons, and no bound states form.
We are therefore led to a model with a fermionic DM particle of mass $m_\chi$ interacting with nucleons through the exchange of a mediating scalar $\phi$ of mass $m_\phi$. The interaction Lagrangian is taken to be
\bea
{\cal L}_{int}= - \,g_\chi\,\phi\,\overline{\chi}\chi\, -\, g_N\,\phi\,\overline{N}N\,,
\eea
where $N=p, n$.  For positive $m_\chi$ and $m_\phi$ and $g_\chi g_N <0$, the force between DM and nucleons is repulsive at tree level, and no bound states form (we assume that the loop-level generated attractive force is subdominant).

%For this specific model at hand, o
One also needs to consider additional constraints on the DM  self-interacting strength from halo shapes and merging galaxy clusters such as the Bullet Clusters. This constraint sets~\cite{Randall:2007ph, Feng:2010gw}
\bea
{\sigma_{\chi\chi}\over m_\chi} \ltap 3~\mbox{GeV}^{-3}\,,
\label{eq:bulletconstraint}
\eea
which also ensures that the DM does not self-scatter sufficiently many times to
form bound states, even though the force between them is attractive.

%There is also a constraint on the couplings to nucleons.
Since the scalar mediates
an attractive force between nucleons, they can in principle form bound states as well.
This is more complicated to analyze because there is already a force between nucleons
due to pion exchange which cannot be calculated precisely. We will therefore require
the new scalar force to be small compared to two-pion exchange; we conservatively
take $m_\phi \sim m_\pi$ and $g_{N} \ltap 0.3\,g_{\pi NN}$ which would lead to small corrections to the nucleon potential~\cite{Beane:2000fx}.

%In terms of the mediator mass and coupling, we have the DM self-scattering cross section as
%\bea
%\sigma_{\chi\chi} = \frac{g_{\chi}^4}{4\pi}\,\frac{m_\chi^2}{m_\phi^4}\,,
%\eea
%where the momenta of non-relativistic DM are below $m_\phi$ and neglected. Similarly, the scattering %cross section $\sigma_{\chi N}$ for non-relativistic DM is calculated to be
%\bea
%\sigma_{\chi p} = \frac{g_{\chi}^2 \,g_{N}^2}{\pi}\,\frac{\mu_{\chi p}^2}{m_\phi^4}\,,
%\eea
%with $\mu_{\chi p}$ as the reduced mass of the DM and proton system.
 For the DM mass equal or below the proton mass and from the constraint in Eq.~(\ref{eq:bulletconstraint}), we find
\bea
\sigma_{\chi p} \ltap \frac{4\,g_N^2}{g_\chi^2}\,\frac{m_\chi}{1~\mbox{GeV}}\times 10^{-27}~\mbox{cm}^2\,,
\eea
where to calculate $\sigma_{\chi \chi}$  and $\sigma_{\chi p}$  we have neglected the non-relativistic DM momenta  which are below $m_\phi$. This constraint is model-dependent; for $g_N \ltap \,g_\chi$ and $m_\chi < 0.2$~GeV one has even a stronger constraint than the existing ones in Fig.~\ref{fig:constraint} from the merging galaxy cluster constraints.

%We can combine these constraints
%$m_a\sim m_\pi$ and $g_{N}=0.1g_{\pi NN}$
%to get
%\bea
%{\sigma_{N\chi}}={1\over \pi}{g_N^2g^2_\chi\over m_a^4}\mu^2<
%\sqrt{12\over \pi}{(0.1g_{\pi NN})^2\over m_\pi^2}\sqrt{\mu^4\over m_\chi GeV^3}
%{3\over \pi}{0.1g_{\pi NN}\over m_\pi^2}\mu^2\sqrt{4\pi/M_\chi GeV^3}
%\eea
%which is the new constraint on the parameter space.

{\it Summary}: As can be seen from Fig.~\ref{fig:constraint}, these combined constraints exclude SIMPs above 2 GeV (cf. \cite{Mack:2007xj}). On the other hand, for  a low mass $m_\chi \ltap 1$ GeV, the constraints become much weaker due to the low-energy threshold limitation of direct detection experiments. After imposing all constraints, we still have a large region of parameter space allowed for SIMPs below 1 GeV with a scattering cross section as large as $10^{-25}$ cm$^2$.

%%%%%%%%%%%%%%%%%%%%%%%%%%%%%%%
{\it{\textbf{Collider Searches for SIMPs.}}}
%%%%%%%%%%%%%%%%%%%%%%%%%%%%%%%
The high-energy colliders do not have the limitation of direct detection experiments on searching for light DM particles. We now explore the possibility of seeing SIMPs at colliders and discuss their spectacular collider signatures.

The first point to note is that a strongly interacting DM particle, if produced at colliders, may not be invisible to detectors. Indeed, if the DM has interaction cross sections similar to hadrons, it would deposit energy in calorimeters, in a manner similar to neutrons and other neutral hadrons. More precisely, the signal depends on the scattering length $L_\chi$, which is inversely proportional to the scattering cross section, of DM inside a detector. If, as is usually the case for WIMPs, $L_\chi$ is significantly larger than the calorimeter size, the DM leaves the collider without any trace and behaves as missing energy. However,  for SIMPs with a large scattering cross section, $L_\chi$ could be smaller than the calorimeter size. The DM will then deposit most or all of its energy in the calorimeters, especially in the hadron calorimeter (HCAL) because it contains heavier elements than the electromagnetic calorimeter (ECAL). Therefore, we anticipate that the SIMP behaves like a jet at colliders.

For a fast moving DM, its inelastic scattering cross section with nucleons (usually much larger than the elastic scattering cross section) determines the scattering length $L_\chi$. We can relate it to the scattering length of neutrons inside materials via
\bea
L_\chi=L_n {\sigma^{\rm inela}_{\chi p}\over \sigma^{\rm inela}_{n p}} \,.
\eea
A typical nucleon-nucleon scattering cross section is about
$40\,\mbox{mb} = 4 \times 10^{-26}\,\mbox{cm}^2$. The bounds on the DM-nucleon
{\it elastic} scattering cross sections, as noted above, are about $10^{-25}~\mbox{cm}^2$, 
and the bounds should be much weaker for DM-nucleon inelastic scattering. It is therefore indeed possible to have DM scattering lengths of similar order to hadronic scattering lengths. 
In fact, since a typical hadron scatters about 10 times inside the HCAL, the DM may appear as a jet even if its cross section is somewhat smaller than hadronic cross sections. 
For $10^{-26}~\mbox{cm}^2 \ltap \sigma^{\rm inela}_{\chi p}$, the SIMPs will appear as jets.

For the rest of this paper, we will focus on the region where the scattering length of the DM particles is much smaller than the calorimeter size, so that 
all the DM energy is deposited inside the detectors~\footnote{The case with an intermediate scattering length is also interesting. One may have two unbalanced trackless jets with missing energy aligning with those two jet momenta. With the help of an additional initial state radiation jet, one may distinguish this signature from backgrounds.}. For pair-produced DM particles, we therefore have a dijet signature. Furthermore, since DM is neutral, the DM jets should have no tracks and less electromagnetic activities in the ECAL and hence are {\it trackless jets}. Exactly because of this property, the DM jets can be distinguished from QCD jets.

\begin{figure}[!]
\begin{center}
\vspace{0.5cm}
\includegraphics[width=0.48\textwidth]{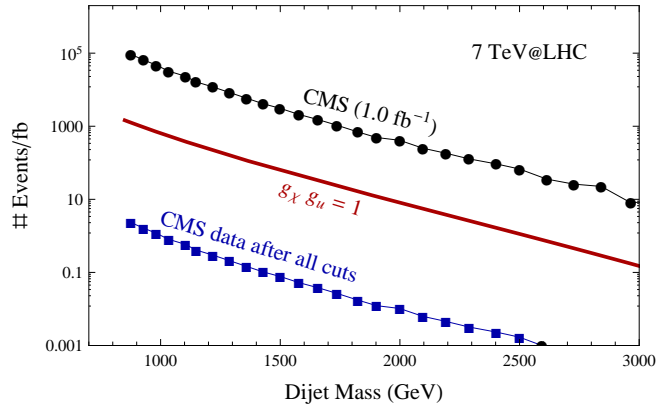} \hspace{2mm} %\\
\caption{The DM jet signal production cross section for $g_\chi g_u =1$ for the solid and red line. The measured dijet invariant mass spectrum is shown in the circle and black points~\cite{Chatrchyan:2011ns}. The squared and blue points are dijet invariant mass spectrum after applying the no-track and less-EMF cut on both jets.
}
\label{fig:production}
\end{center}
\end{figure}

To determine the feasibility of detecting these events, we first need to
know the number of such events. This depends on further details
of the model, in particular the interaction of the mediator with quarks and gluons
(rather than nucleons).  Using the matrix element of quarks inside nucleons, 
one can relate the mediator coupling to quarks $g_q$ to the couplings to nucleons $g_N$. For example, one has $g_{N} = 8.22 g_u$ for $N=p$~\cite{Belanger:2008sj}. In terms of the coupling to quarks, we have the production cross section of $\sigma( u \bar{u} \rightarrow \bar{\chi} \chi) = g_\chi^2 g^2_u /(12\pi \hat{s})$ (where we have neglected the light DM mass). Using the MSTW parton distribution function~\cite{Martin:2009iq}, we have the signal production cross section at the 7 TeV LHC shown in the solid (red) line of Fig.~\ref{fig:production}. 

The background, which comes from QCD dijet production,
has been measured~\cite{Chatrchyan:2011ns} and found to be in  good agreement 
with the SM prediction.  We show the observed data in the circle dots (blue) of
 Fig.~\ref{fig:production} as an estimate of the background.

We have two main handles to reduce the QCD backgrounds. First, we require that there are no tracks in the tracker system. The charged particle multiplicity has been measured by the CMS collaboration at 0.9, 2.36 and 7 TeV~\cite{Khachatryan:2010nk} and approximately follows the Koba-Nielsen-Olesen (KNO) scaling $P(n) =\langle n \rangle^{-1} e^{-n/\langle n \rangle}$~\cite{Koba:1972ng}. The average charged-particle number is $\langle n \rangle \approx 20$~\cite{Khachatryan:2010nk}, and hence approximately 5\% of QCD jets have no charged particles in them. Requiring two trackless jets therefore reduces the background by a factor of 400.

We also require a small electromagnetic fraction (EMF) for each jet. For QCD jets, the EMF 
(coming from $\pi^0 \rightarrow \gamma\gamma$ and radiations of charged hadrons)
is typically peaked at 50\%~\cite{AtlasEMF} while for DM jets the EMF is proportional to the scattering length of DM in the ECAL over that in the HCAL and is around 10\%. We can hence impose a cut that the EMF should be less than 20\%. This further reduces the background by a factor of about 100. We note that there could be a correlation between the no-track cut and the EMF cut and the actual reduction on the QCD backgrounds requires detailed detector simulations.

Applying those two cuts, the QCD backgrounds can be reduced by a factor of around 40000; the reduced backgrounds are shown in the squared points (blue) in Fig.~\ref{fig:production}. 
We see that these cuts reduce the background to well
below the DM production rate. Hence such DM jets can be discovered even at the early LHC.

%%%%%%%%%%%%%%%%%%%%%%%%%%%%%%%
{\it{\textbf{Conclusions.}}}
%%%%%%%%%%%%%%%%%%%%%%%%%%%%%%%
We have argued that light DM particles which have strong interactions
with  the SM particles are not excluded by current astrophysical
constraints. These particles can even have interaction cross sections close to
hadronic interactions. We have also shown that these DM particles can have unique signatures at hadron colliders. Unlike WIMPs, these DM particles do not
manifest themselves as missing energy; instead there are jets associated with
them. We have proposed a new search strategy for SIMPs by looking for a signal of two trackless jets. We have shown that suitable cuts can plausibly reduce the QCD dijet background so that the DM jets may be discovered even in the early data of the LHC.

\vspace*{0.06in}
We thank G. Landsberg, N. Makovec, M. Peskin, D. Whiteson, and D. Zerwas for useful discussions, and especially C. S. Hill for encouragement on this project. This material is based upon work supported in part by the NSF under Grant No. 1066293 and the hospitality of the Aspen Center for Physics. The work of AR  is supported in part by NSF grant PHY-0709742. SLAC is operated by Stanford University for the US Department of Energy under contract DE-AC02-76SF00515.

\end{document}